\documentclass[twocolumn]{aastex631}
\usepackage{booktabs}
\usepackage{footmisc}

\newcommand{\uat}[2]{#1 (#2)}
\newcommand{\kms}{km\,s$^{-1}$}

\newcommand{\cont}{1.3\,mm continuum}
\newcommand{\cto}{$^{12}$CO\,(2 $-$ 1)}
\newcommand{\sio}{SiO\,(5 $-$ 4)}
\newcommand{\ceo}{C$^{18}$O}
\newcommand{\ntdp}{N$_2$D$^+$}
\newcommand{\dcop}{DCO$^+$}
\newcommand{\dcn}{DCN}
\newcommand{\choh}{CH$_3$OH}
\newcommand{\chohlower}{CH$_3$OH\,($4_2 - 3_1$)}
\newcommand{\chohhigher}{CH$_3$OH\,(10$_2$ $-$ 9$_3$)}
\newcommand{\htco}{H$_2$CO}
\newcommand{\htcoi}{H$_2$CO\,3(0, 3) $-$ 2(0, 2)}
\newcommand{\htcoii}{H$_2$CO\,3(2, 2) $-$ 2(2, 1)}
\newcommand{\htcoiii}{H$_2$CO\,3(2, 1) $-$ 2(2, 0)}
\newcommand{\htdp}{H$_2$D$^+$}
\newcommand{\chtdp}{CH$_2$D$^+$}

\begin{document}

\title{ALMASOP: Inner-Envelope Structures of Protostars Driving Nascent Jets}

\author[0000-0002-2338-4583]{Somnath Dutta}
\email{sdutta@asiaa.sinica.edu.tw}
\affiliation{Academia Sinica Institute of Astronomy and Astrophysics, Roosevelt Rd, Taipei 106319, Taiwan, R.O.C.}

\author[0000-0002-3024-5864]{Chin-Fei Lee}
\affiliation{Academia Sinica Institute of Astronomy and Astrophysics, Roosevelt Rd, Taipei 106319, Taiwan, R.O.C.}
\author[0000-0001-9304-7884]{Naomi Hirano}
\affiliation{Academia Sinica Institute of Astronomy and Astrophysics, Roosevelt Rd, Taipei 106319, Taiwan, R.O.C.}

\author[0000-0002-6773-459X]{Doug Johnstone}
\affiliation{\textnormal{NRC Herzberg Astronomy and Astrophysics, 5071 West Saanich Rd, Victoria, BC, V9E 2E7, Canada }}
\affiliation{\textnormal{Department of Physics and Astronomy, University of Victoria, Victoria, BC, V8P 5C2, Canada}}

\author[0000-0003-2412-7092]{Kee-Tae Kim}
\affil{Korea Astronomy and Space Science Institute (KASI), 776 Daedeokdae-ro, Yuseong-gu, Daejeon 34055, Republic of Korea}
\affil{University of Science and Technology, Korea (UST), 217 Gajeong-ro, Yuseong-gu, Daejeon 34113, Republic of Korea}

\author[0000-0002-4336-0730]{Yi-Jehng Kuan}
\affiliation{Department of Earth Sciences, National Taiwan Normal University, Taipei, Taiwan, R.O.C.}
\affiliation{Academia Sinica Institute of Astronomy and Astrophysics, Roosevelt Rd, Taipei 106319, Taiwan, R.O.C.}

\author{James Di Francesco}
\affiliation{\textnormal{NRC Herzberg Astronomy and Astrophysics, 5071 West Saanich Rd, Victoria, BC, V9E 2E7, Canada }}
\affiliation{\textnormal{Department of Physics and Astronomy, University of Victoria, Victoria, BC, V8P 5C2, Canada}}

\author[0000-0002-8149-8546]{Ken'ichi Tatematsu}
\affil{Nobeyama Radio Observatory, National Astronomical Observatory of Japan, 
National Institutes of Natural Sciences, 
462-2 Nobeyama, Minamimaki, Minamisaku, Nagano 384-1305, Japan}
\affiliation{Department of Astronomical Science,
SOKENDAI (The Graduate University for Advanced Studies),
2-21-1 Osawa, Mitaka, Tokyo 181-8588, Japan}

\author[0000-0002-5809-4834]{Mika Juvela}
\affiliation{Department of Physics, P.O.Box 64, FI-00014, University of Helsinki, Finland}

\author{Chang Won Lee}
\affil{Korea Astronomy and Space Science Institute (KASI), 776 Daedeokdae-ro, Yuseong-gu, Daejeon 34055, Republic of Korea}
\affil{University of Science and Technology, Korea (UST), 217 Gajeong-ro, Yuseong-gu, Daejeon 34113, Republic of Korea}

\author{Alessio Traficante}
\affil{IAPS-INAF, via Fosso del Cavaliere 100, I-00133, Rome, Italy}

\author{Vivien Huei-Ru Chen}
\affiliation{Institute of Astronomy and Department of Physics, , National Tsing Hua University, Hsinchu, 300044, Taiwan}

\author{Manash Ranjan Samal}
\affiliation{Physical Research Laboratory, Navrangpura, Ahmedabad, Gujarat 380009, India}

\author{David Eden}
\affiliation{Armagh Observatory and Planetarium, College Hill, Armagh, BY61 9DB, United Kingdom}

\author[0000-0002-4393-3463]{Dipen Sahu}
\affiliation{Physical Research Laboratory, Navrangpura, Ahmedabad, Gujarat 380009, India}
\affiliation{Academia Sinica Institute of Astronomy and Astrophysics, Roosevelt Rd, Taipei 106319, Taiwan, R.O.C.}

\author[0000-0002-1369-1563]{Shih-Ying Hsu}
\affiliation{Academia Sinica Institute of Astronomy and Astrophysics, Roosevelt Rd, Taipei 106319, Taiwan, R.O.C.}

\author[0000-0002-5286-2564]{Tie Liu}
\affiliation{Shanghai Astronomical Observatory, Chinese Academy of Sciences, 80 Nandan Road, Shanghai 200030, China}

\author[0000-0003-4603-7119]{Sheng-Yuan Liu}
\affiliation{Academia Sinica Institute of Astronomy and Astrophysics, Roosevelt Rd, Taipei 106319, Taiwan, R.O.C.}

\begin{abstract}
Protostellar jets provide valuable insight into the evolutionary stage and formation history of star-forming cores in their earliest phases. We investigated the inner envelope structures of three extremely young protostars, selected for having the shortest dynamical timescales in their outflows and jets. Our analysis is based on Atacama Large Millimeter/submillimeter Array (ALMA) observations of the \ntdp{}, \dcop{}, \dcn{}, \ceo{}, \choh{}, and \htco{} lines, along with 1.3 mm continuum data, obtained at two spatial resolutions of $\sim$500 AU and 150 AU. By examining molecular depletion and sublimation patterns, emission extents at core-scale and outflow rotational temperatures, we assessed the relative evolutionary stages of the three sources. In G208.68-19.20N1, the absence of \ntdp{} toward the core—despite a semi-ring-like distribution—and the presence of bright \dcn{} and \dcop{} emission co-spatial with \ceo{} indicate a warmer envelope,  possibly suggesting a more advanced evolutionary state. In contrast, G208.68-19.20N3 shows no dense central structures in \ceo{}, \dcn{}, \dcop{}, or \ntdp{}, with emission instead appearing scattered around the continuum and along large-scale filaments, consistent with a likely younger stage than G208.68-19.20N1. The third source, G215.87-17.62M, exhibits compact \ceo{} emission at the continuum peak, but spatially extended \ntdp{}, \dcn{}, and \dcop{} along the continuum, pointing to a cooler envelope and likely the youngest stage among the three. This comparative analysis across three protostars demonstrates the effectiveness of molecular tracers for evolutionary staging, though variations in luminosity or accretion may also shape chemical morphologies. These results highlight the promise of broader surveys for advancing our understanding of early protostellar evolution.
\end{abstract}

\keywords{\uat{Protostars}{1302} --- \uat{Stellar jets}{1607} ---  \uat{Submillimeter astronomy}{1647} --- \uat{Astrochemistry}{75} --- \uat{Circumstellar envelopes}{237} --- \uat{Star formation}{1569}}

\section{Introduction} \label{sec:intro}
The internal structure of the envelope in the earliest phases of protostellar evolution offers a unique window into the initial environmental conditions shaping core development. During this stage, increasing central luminosity raises core temperatures, heating the surrounding material and driving chemical evolution. A detailed investigation of the envelope's chemical structure at this phase is therefore crucial for understanding the processes governing early protostellar evolution.

Different molecular species in protostellar cores can be correlated through their depletion, sublimation, temperature dependence, and spatial distribution of emission. For example, deuterated species like \ntdp{} show enhanced fractional abundance in cores on the verge of stellar collapse, when the gas temperature is $\sim 10$~K \citep[e.g.,][]{1999ApJ...523L.165C,2015A&A...578A..55S,2020ApJS..249...33K,2022ApJ...939..102L}. As the core evolves and the surrounding gas is heated above the CO sublimation temperature ($\sim 25$~K), molecular species synthesized from sublimated CO become detectable. Gas-phase CO can produce \dcop{} by destroying \htdp{} or \ntdp{} via  
\begin{equation}
\mathrm{H_2D^+} + \mathrm{CO} \rightarrow \mathrm{DCO^+} + \mathrm{H_2}
\end{equation}
and  
\begin{equation}
\mathrm{N_2D^+} + \mathrm{CO} \rightarrow \mathrm{DCO^+} + \mathrm{N_2},
\end{equation}
respectively \citep[e.g.,][]{1999ApJ...523L.165C,2004ApJ...617..360L}.  
\dcn{} can also form in warmer gas, where thermal desorption releases it from icy grain mantles into the gas phase. In outflows or accretion shocks (T~$>$~100~K), neutral--neutral reactions such as  
\begin{equation}
\mathrm{CN} + \mathrm{HD} \rightarrow \mathrm{DCN} + \mathrm{H}
\end{equation}
become important for \dcn{} production \citep[e.g.,][]{2006A&A...449..621F}.  
Consequently, \ntdp{} depletion is expected in regions of high CO abundance, accompanied by enhanced \dcop{} and \dcn{} emission \citep[e.g.,][]{1995A&A...303..541J,2009A&A...508..737P}. Thus, the detection, depletion, and sublimation of specific molecular species diagnose local heating and the evolutionary status of protostellar cores, and Atacama Large Millimeter/submillimeter Array (ALMA) observations of dense gas tracers at submillimeter wavelengths are crucial for revealing the morphology, kinematics, and chemical composition of their high-density envelopes.

Protostellar jets are launched at the onset of collapse and serve as signposts of the accretion process \citep[e.g.,][]{2001ARA&A..39..403R,2014prpl.conf..451F,2016ARA&A..54..491B,2020A&ARv..28....1L}. In the earliest stages, the high density of circumstellar material confines the jets within the envelope \citep[e.g.,][]{2007prpl.conf..245A}, resulting in smaller spatial extents and younger dynamical ages \citep[e.g.,][]{1996A&A...311..858B,2024AJ....167...72D}. Consequently, protostellar systems hosting the youngest jets provide unique laboratories for probing freshly formed molecules in the immediate vicinity of the protostar, offering insights into the earliest stages of chemical and dynamical evolution.

\begin{deluxetable*}{lcccc}
\tablecaption{Observed Parameters of the Protostars \label{tab:summary_parameters}}
\tablewidth{0pt}
\tablehead{
  \colhead{Parameter} & \colhead{G208N1} & \colhead{G208N3} & \colhead{G215M} & \colhead{References}
}
\startdata
$T_{\rm bol}$ (K) & $38\pm13$ & \nodata & \nodata & \tablenotemark{1}, \tablenotemark{2}, \tablenotemark{3} \\
$L_{\rm bol}$ (L$_\odot$) & $36.7\pm14.5$ & \nodata & \nodata & \tablenotemark{1}, \tablenotemark{2}, \tablenotemark{3} \\
$M_{\rm env}$ (M$_\odot$) & $8.26\pm1.28$ & $2.08\pm0.34$ & $0.30\pm0.07$ & \tablenotemark{4} \\
$R_{\rm env@500AU}$ (AU) & 4000--5000 & 1200--2000 & 1000--2400 & This study \\
$\dot{M}_{\rm jet}$ (10$^{-6}$ M$_\odot$\,yr$^{-1}$) & 2.5 & 1.3 & 0.32 & \tablenotemark{4}\\
$V_{\rm jet}$ (km\,s$^{-1}$) & $102^{+24}_{-35}$ & $61^{+41}_{-15}$ & $60^{+28}_{-13}$ & \tablenotemark{4}\\
$F_{\rm CO}$ (10$^{-6}$ km\,s$^{-1}$\,yr$^{-1}$) & 2.54 & 1.70 & 1.15 & \tablenotemark{4}\\
$\tau_{\rm dyn}$ (yr) & 47$^{+38}_{-10}$ & 228$^{+160}_{-110}$ & $292^{+133}_{-108}$ & \tablenotemark{4} \\
$T_{\rm rot}$ (K) & 75--164 & 110--193 & 50--54 & This study \\
\enddata
\tablenotetext{1}{\citet[][]{2016ApJS..224....5F}}
\tablenotetext{2}{\citet[][]{2020ApJS..251...20D}}
\tablenotetext{3}{\citet[][]{2020ApJ...890..130T}}
\tablenotetext{4}{\citet[][]{2024AJ....167...72D}}
\end{deluxetable*}

In this paper, we investigate the envelopes of three protostars — G208.68-19.20N1 (G208N1, also known as HOPS\ 87), G208.68-19.20N3 (G208N3), and G215.87-17.62M (G215M) — which exhibit molecular SiO and CO emission in their high-density jets, indicative of young jet activity. These sources are located within the Orion molecular cloud complex. The morphology of their jets has been examined in \citet{2024AJ....167...72D} through SiO and CO emission using ALMA observations, as part of the ALMASOP project \citep[ALMA Survey of Orion Planck Galactic Cold Clumps;][]{2020ApJS..251...20D}.  Table \ref{tab:summary_parameters} outlines the main characteristics of the protostars. G208N1 is a Class\,0 protostar with $T_{\rm bol} = 38 \pm 13$~K and $L_{\rm bol} = 36.7 \pm 14.5\,L_{\odot}$ \citep[][]{2016ApJS..224....5F, 2020ApJ...890..130T, 2020ApJS..251...20D}, and from 1.3~mm continuum observations at $\sim$2000~AU resolution, its envelope mass was derived to be $M_{\rm env} \approx 8.26 \pm 1.28\,M_{\odot}$. G208N3, located $\sim$16\arcsec\ from a close binary system (N3B and N3C) \citep[][]{2020ApJS..251...20D}, remains poorly characterized in earlier studies with no estimates of $T_{\rm bol}$ or $L_{\rm bol}$ reported; however, continuum data yield an envelope mass of $M_{\rm env} \sim 2.08 \pm 0.34\,M_\odot$. G215M is the least studied of the three, with no $T_{\rm bol}$ or $L_{\rm bol}$ estimates available, and its envelope mass is measured as $M_{\rm env} \sim 0.30 \pm 0.07\,M_\odot$.  Based on the spatial extents and deprojected velocities of the three jets, as determined by \citet{2024AJ....167...72D}, these protostars appear to have  extremely young dynamical ages ($<$500~yr) and moderately high jet mass-loss rates of $(0.3$--$2.5) \times 10^{-6}$~M$_\odot$~yr$^{-1}$. These characteristics make them excellent examples of extremely young protostellar systems that have only recently begun their accretion-driven outflow activity. Studying such objects provides valuable insight into the physical and chemical conditions at the earliest stages of protostellar evolution. The structure of the paper is as follows: Section 2 describes the data used in this study; Section 3 presents the observational results; Section 4 discusses the evolutionary status of the sources; and Section 5 summarizes our findings and presents the main conclusions.

\section{Observations}\label{sec:observations}
\subsection{Data from ALMASOP}
We have utilized ALMA archive data observed as a part of the ALMA Survey of Orion Planck cold clumps (ALMASOP; Project ID:2018.1.00302.S) in Band\,6 \citep[see][for more details on ALMASOP]{2020ApJS..251...20D}. In this paper, we make use of \ceo{}, \ntdp{}, \dcn{}, \dcop{}, \choh{}, and \htco{}. For \choh{} we have studied two transitions at \chohlower{} and \chohhigher{}. For \htco{}, we have created maps at three transitions \htcoi{}, \htcoii{} and \htcoiii{}. Following \citet[][]{2024AJ....167...72D},  we reproduced  CO (2$-$1) and SiO (5$-$4) maps. Although additional lines are detected toward some individual protostars, we selected the present set of transitions because they are the most common molecular tracers in young protostars, and are well suited to probe both the envelope and outflow components. They are detected in the majority of our studied sources. While a few lines are missing in certain objects, these absences are themselves scientifically meaningful and provide additional insight into the diversity of physical and chemical conditions across our sample. The spectral lines used in this work, along with their rest frequency ($\nu$), upper-state energy ($E_\mathrm{up}$), and logarithmic Einstein coefficient ($A_\mathrm{ul}$), are listed in Table~\ref{tab:Spectral_lines_used_in_this_work}. The spectroscopic parameters were adopted from the laboratory measurements reported in the literature \citep[e.g.,][]{2000JMoSp.200..143M,2001A&A...370L..49M,2016JMoSp.327...95E,2017JMoSp.331...28M}, as compiled in the Cologne Database for Molecular Spectroscopy \citep[CDMS;][]{2005JMoSt.742..215M} and made available through the Splatalogue database\footnote{\url{https://splatalogue.online/}}. Data calibration has been performed with the standard pipeline in CASA\ 5.5 \citep[]{2007ASPC..376..127M}.  We have generated two sets of line cubes for all transitions and continuum emission: (i) by combining all visibilities observed with 12\,m C43-5 (TM1), 12\,m C43-2 (TM2), and 7 m ACA \citep[see][for more details]{{2020ApJS..251...20D}}, where the synthesized beam size is $\sim$ 0$\farcs$41 $\times$ 0$\farcs$34, and (ii) with a UV taper of $\sim$ 1$\farcs$0 by combining all visibilities (TM1+TM2+ACA), which leads to synthesize beam size of $\sim$ 1$\farcs$5 $\times$ 1$\farcs$2. Throughout this paper, we refer to the first case as high-resolution ($\sim$ 150 AU at Orion distance)  and the latter as low-resolution ($\sim$ 500 AU at Orion distance). The combined maps of TM1+TM2+ACA have a maximum recoverable scale (MRS) of $\sim 14\arcsec$. In all cases, we have applied a robust weighting factor of R$_w$ = $+$2.0 (natural weighting) to enhance sensitivity to extended emission. The line cube channels were binned to 2~km~s$^{-1}$ to improve sensitivity. The continuum peak positions at high resolution, together with the continuum sensitivities and the mean per-channel sensitivity of the line cubes at both high and low resolution, are listed in Table \ref{tab:sources_position_ContSensitivity}. The ALMA flux calibration uncertainty is estimated to be $\sim 10\%$, consistent with standard values for Band 6 observations.

\begin{table*}[t]
\centering
\caption{Spectral lines used in this work}
\label{tab:Spectral_lines_used_in_this_work}
\begin{tabular}{l l r c c}
\hline
Molecule & Transition & $\nu$ (GHz) & $E_\mathrm{up}$ (K) & log$_{10}$($A_\mathrm{ul}$) (s$^{-1}$)\\
         &            &             &                     &                                       \\
\hline
\hline
C$^{18}$O & $2\!-\!1$ & 219.5603541 & 15.81 & $-6.22$  \\
N$_2$D$^+$ & $3\!-\!2$ & 231.3199064 & 22.20 & $-4.05$  \\
DCN & $3\!-\!2$ & 217.2383000 & 20.85 & $-3.34$  \\
DCO$^+$ & $3\!-\!2$ & 216.1125705 & 20.74 & $-2.62$  \\
CH$_3$OH & $4_{2}\!-\!3_{1}$ & 218.4400630 & 45.46 & $-4.33$  \\
CH$_3$OH & $10_{2}\!-\!9_{3}$ & 231.2811100 & 165.35 & $-4.74$  \\
H$_2$CO & $3_{0,3}\!-\!2_{0,2}$ & 218.2221920 & 20.96 & $-3.55$  \\
H$_2$CO & $3_{2,2}\!-\!2_{2,1}$ & 218.4756320 & 68.09 & $-3.80$  \\
H$_2$CO & $3_{2,1}\!-\!2_{2,0}$ & 218.7600660 & 68.11 & $-3.80$  \\
$^{12}$CO & $2\!-\!1$ & 230.5380000 & 16.60 & $-6.16$  \\
SiO & $5\!-\!4$  & 217.1049800 & 31.26 & $-3.28$  \\
\hline
\end{tabular}
\end{table*}

\subsection{Data from the ALMA Archive}
We have retrieved ALMA band 6 data observed in Cycle-3 (\#2015.1.00341.S; PI; S. Takahashi) at high spectral resolution. Among the three objects studied here, two were included in this project G208.68-19.20N1 as MMS6 and G208.68-19.20N3 as MMS2. The SPW, which includes \ntdp{} line, was set to the high spectral resolution mode with a 35.28 kHz resolution and the corresponding velocity resolution was $\sim$ 0.046 \kms{}. The \ntdp{} images were created using the data observed with the ACA 7m array and C36-1 configuration of the 12\,m array. The visibility data were calibrated using CASA. 

The calibrated visibility data from the two configurations were combined and imaged with the TCLEAN task in CASA\ 5.5, after continuum subtraction, to produce a line image cube. We applied Briggs weighting with a robust parameter of +0.5. Further details of the data reduction are provided in \citet[][]{2024ApJ...961..123H}. The final \ntdp{} image cubes have a synthesized beam of $\sim$1$\farcs$5 $\times$ 0$\farcs$9, a velocity resolution of $\sim$0.1 \kms{}, and a per-channel sensitivity of $\sim$ 33 mJy beam$^{-1}$. These archival observations have a similar resolution and recoverable scale to the ALMASOP low-resolution 1\arcsec\ UV-tapered maps, although they have a significantly higher velocity resolution ($\sim 20$ times) compared with ALMASOP. The absolute flux calibration uncertainty of these  Band\,6 observations is $\sim $10\%, based on the ALMA calibrator measurements.



\begin{table*}[t]
\centering
\caption{Continuum Peak Positions and Continuum/Spectral Sensitivities (ALMASOP)}
\label{tab:sources_position_ContSensitivity}
\begin{tabular}{l c c c c c c c}
\toprule
Source & RA (h:m:s) & Dec (d:m:s) &
\multicolumn{2}{c}{Continuum rms (mJy beam$^{-1}$)} &
\multicolumn{2}{c}{Line rms/channel (mJy beam$^{-1}$)} &
Vel Res (km s$^{-1}$) \\
\cmidrule(lr){4-5} \cmidrule(lr){6-8}
 & & & @150 AU & @500 AU & @150 AU & @500 AU & \\
\midrule
G208N1 & 05:35:23.42 & $-$05:01:30.60 & 0.20 & 1.50 & 3.5 & 5.8 & 2 \\
G208N3 & 05:35:18.06 & $-$05:00:18.19 & 0.20 & 0.14 & 3.4 & 5.5 & 2 \\
G215M  & 05:53:32.52 & $-$10:25:08.18 & 0.04 & 0.04 & 3.2 & 5.2 & 2 \\
\bottomrule
\end{tabular}
\end{table*}

\begin{figure*}
    \fig{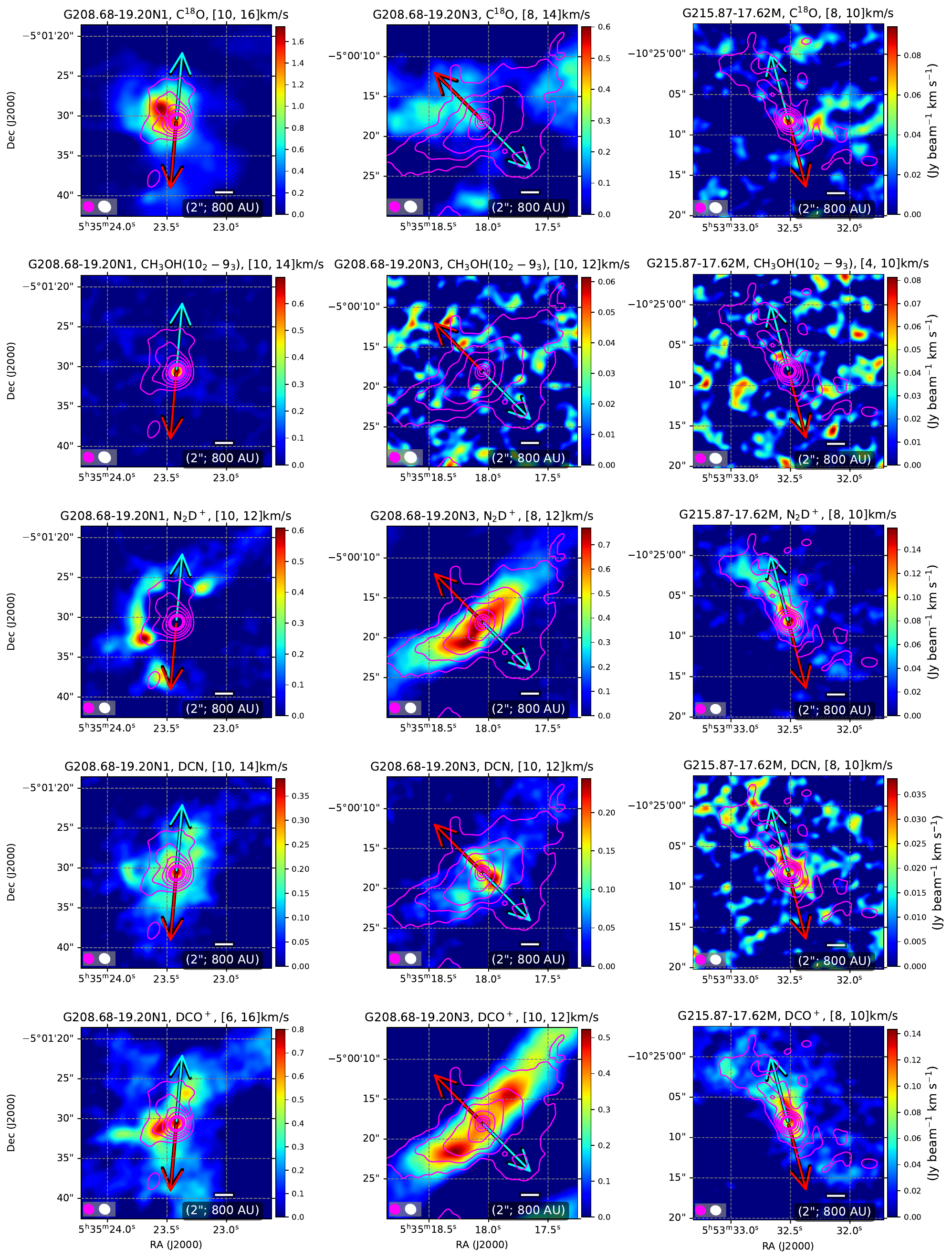}{0.88\textwidth}{}
    \caption{Low–angular resolution maps at a spatial scale of $\sim$500 AU are shown for different molecular tracers, primarily tracing the envelope structures. The three columns correspond to three sources: G208.68–19.20N1 (G208N1), G208.68–19.20N3 (G208N3), and G215.87–17.62M (G215M), respectively. The background images display molecular emission maps integrated over the specified velocity ranges indicated at the top of each panel. Magenta contours, plotted at the same angular resolution as the molecular emission, represent the continuum emission at levels of (6, 18, 36, 72, 144, 200, 250) $\times$ $\sigma$, where $\sigma$ is 1.5, 0.14, and 0.04 mJy beam$^{-1}$ km s$^{-1}$ for G208N1, G208N3, and G215M, respectively. Jet axis directions are indicated with red arrows for redshifted emission and cyan arrows for blueshifted emission. Beam sizes are shown in the lower left of each panel, with magenta representing the continuum and white representing the molecular emission. A spatial scale bar is provided. 
    }
    \label{fig:EnvelopeEmission_lowresolution}
\end{figure*}

\begin{figure*}
    \centering
    \includegraphics[width=0.9\linewidth]{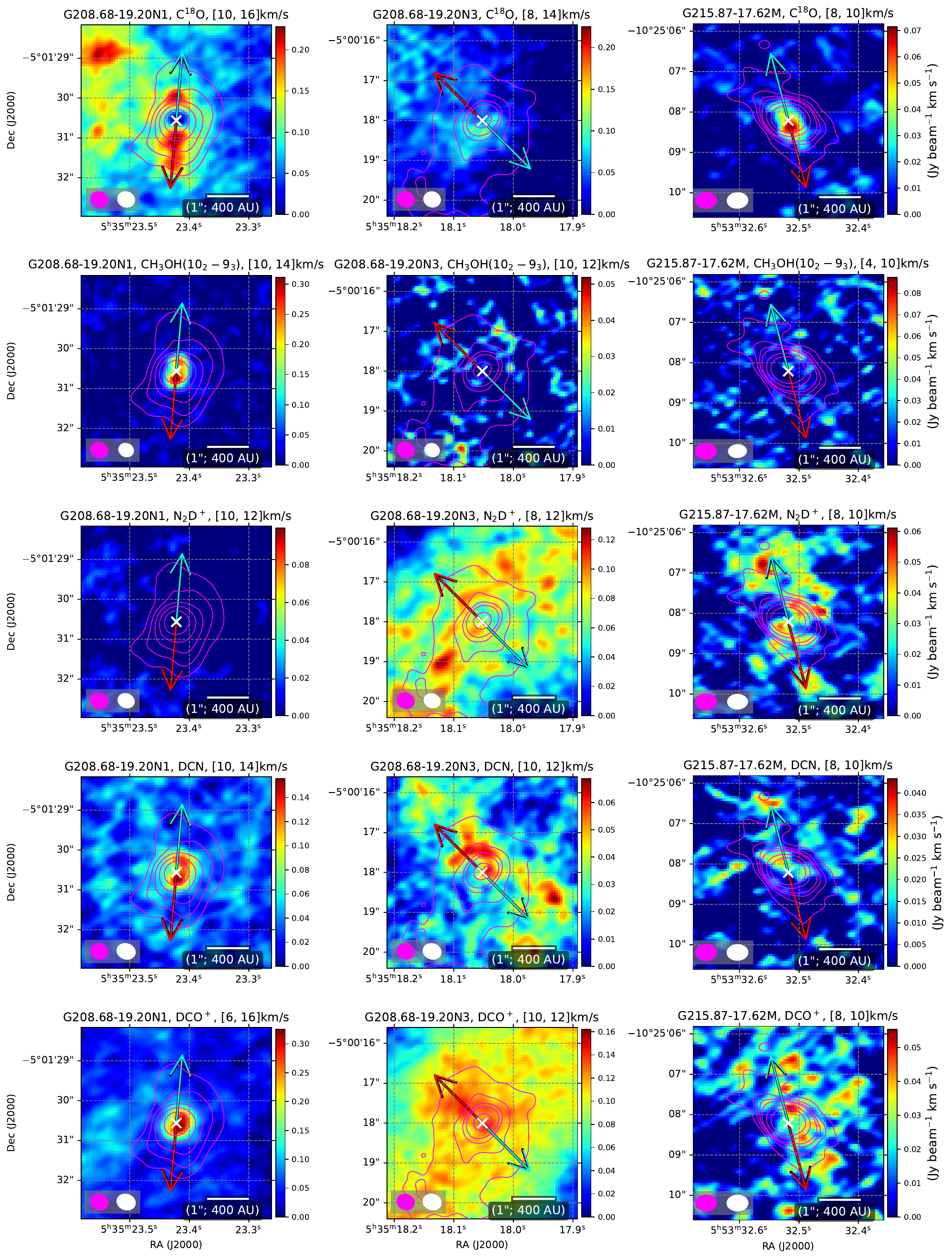}
    \caption{High--angular resolution maps at a spatial scale of $\sim$150~AU are shown. All symbols are the same as in Figure~\ref{fig:EnvelopeEmission_lowresolution}. The continuum contours, plotted at the same angular resolution as the molecular emission, are at (50, 100, 200, 400, 800)$\times\sigma$ ($\sigma = 0.2$~mJy~beam$^{-1}$~km~s$^{-1}$) for G208N1, and (6, 12, 18, 24, 50, 100)$\times\sigma$  for G208N3 ($\sigma = 0.2$~mJy~beam$^{-1}$~km~s$^{-1}$) and  for G215M ($\sigma = 0.04$~mJy~beam$^{-1}$~km~s$^{-1}$). The Continuum peaks are marked with white crosses.
    }
    \label{fig:EnvelopeEmission_highresolution}
\end{figure*}

\begin{table*}[ht]
\centering
\caption{N$_2$D$^+$ Fit Parameters}
\begin{tabular}{lcccc}
\hline
Position & $T_{ex}$ (K) & $\tau$ & $V_{LSR}$ (km/s) & V$_{\rm FWHM}$ (km/s) \\
\hline\hline
& & G208N1 & &\\
\hline
p1 & 12.34$\pm$1.25 & 4.63$\pm$2.14 & 10.95$\pm$0.02 & 0.10$\pm$0.02 \\
p2 & 11.18$\pm$2.07 & 2.92$\pm$2.44 & 11.51$\pm$0.04 & 0.20$\pm$0.05 \\
p3 & 10.29$\pm$1.84 & 3.55$\pm$2.77 & 11.47$\pm$0.03 & 0.11$\pm$0.03 \\
p4 & 10.68$\pm$7.33 & 1.60$\pm$3.26 & 11.31$\pm$0.05 & 0.20$\pm$0.08 \\
p5 & 12.82$\pm$2.08 & 3.17$\pm$1.93 & 10.90$\pm$0.02 & 0.12$\pm$0.02 \\
\hline\hline
& & G208N3 & &\\
\hline
p1 & 21.18$\pm$0.54 & 10.24$\pm$1.33 & 11.11$\pm$0.01 & 0.12$\pm$0.01 \\
\hline
\end{tabular}
\tablecomments{The parameters for position p4 have unusually large uncertainties due to low signal-to-noise in the satellite lines.}
\label{tab:n2dp_fitting_params}
\end{table*}

\section{Results} \label{sec:results}

\subsection{Dense Gas Tracers around the envelope} \label{sec:densegastracers}
 Multiple molecular tracers of the gas surrounding the protostellar envelopes are shown in Figure~\ref{fig:EnvelopeEmission_lowresolution} at a low resolution of $\sim$ 500AU, and in Figure~\ref{fig:EnvelopeEmission_highresolution} at a higher resolution of $\sim$ 150~AU.  Each figure consists of three columns, corresponding to the three target sources. The background images show integrated molecular line maps, with the velocity ranges indicated in each panel. We consider the integrated emission to be detected when the emission in the integrated maps exceeds 3$\sigma$. The maps are produced by integrating over velocity channels where the emission in each channel exceeds the 3$\sigma$ level. Overlaid in magenta are continuum contours at comparable resolution, tracing primarily the dust emission. The jet axes, determined from the CO and SiO emission (Figure~\ref{fig:Appendix_COSiO_outflowJet_highresolution}), are shown as arrows. 
In the following sections, we describe the envelope morphology for each of the three objects, highlighting the contributions of different molecular tracers and dust emission across both spatial scales, from lower to higher resolution.

G208N1 (first column of Figure \ref{fig:EnvelopeEmission_lowresolution} and \ref{fig:EnvelopeEmission_highresolution}) is embedded within a large \ceo{} core at a resolution of $\sim$ 500~AU; although compact in appearance, the \ceo{} peak lies northeast of the continuum peak (Figure \ref{fig:EnvelopeEmission_lowresolution}). A close-up view of the high-resolution ($\sim$ 150 AU) \ceo{} map  shows that the emission near the source is elongated along the outflow axis (PA~$\sim175^\circ$; see Section~\ref{sec:Appendix_outflowjet}), with a central gap coincident with the \cont{} peak (Figure \ref{fig:EnvelopeEmission_highresolution}).   Compact \chohhigher{} emission is detected within the inner region of the \cont{}, and the high-resolution \chohhigher{} map reveals structures likely tracing the jet base.   Notably, \ntdp{} is either absent or significantly depleted near the protostellar core, forming an arc-like boundary from the north through east to south around the continuum peak. The \dcn{} emission is more extended at $\sim$500~AU resolution; in the high-resolution ($\sim$150~AU) maps it becomes concentrated at the continuum peaks and is predominantly distributed near the jet base. This suggests that \dcn{} emission may occupy regions left vacant by the absence of \ntdp{}.   Finally, \dcop{} shows an anticorrelation with \ceo{} on both spatial scales: at low resolution it is relatively extended, while at high resolution it is concentrated mainly at the continuum peak. From a visual inspection of the molecular line and continuum emission maps at $\sim$ 500~AU resolution, we estimate that the envelope has an approximate diameter of 4000–6000 AU.

G208N3 (second column of Figure \ref{fig:EnvelopeEmission_lowresolution} and \ref{fig:EnvelopeEmission_highresolution}) exhibits extended \ceo{} emission in the envelope at both high and low spatial scales.  \chohhigher{} is non-detected around this source. The low-resolution  \ntdp{} map suggests that the core is located within a large filamentary cloud elongated from northwest to southeast. In the high-resolution map, \ntdp{} appears scattered along the filaments; a close inspection of the continuum peak indicates that it is mostly scattered and depleted near the continuum peak.  Low-resolution maps reveal that the bright \dcn{} is elongated along the outflow direction (PA $\sim$ 45$^\circ$, see Section~\ref{sec:Appendix_outflowjet}), from northeast to southwest near the core. The G208N3 core itself is also situated within a large filamentary cloud that extends similarly to the \ntdp{} emission. In the high-resolution \dcn{} map, the emission forms a partial ring-like structure around the continuum peak toward the west and northeast.  At low resolution, \dcop{} shows a filamentary distribution similar to that of \ntdp{}, with three dense cores embedded within the observed \dcop{} filaments; G208N3 is located in one of these dense cores.   In the high-resolution maps, \dcop{} emission is more scattered and peaks away from the continuum center, primarily along the northeastern jet axis. Based on a visual inspection of the molecular line emission and continuum extent at resolution of $\sim$ 500~AU, we estimate that the envelope of G208N3 spans a diameter of $\sim$1200--2000~AU.

G215M (third column of Figures~\ref{fig:EnvelopeEmission_lowresolution} and \ref{fig:EnvelopeEmission_highresolution}) exhibits compact \ceo{} emission centered on the continuum peak in both the low- and high-resolution maps, with a slight elongation along the jet direction (PA~$\sim15^\circ$) that is more prominent in the high-resolution map.  \chohhigher{} is not clearly detected in any of the maps toward the continuum.  \ntdp{} emission is present in both spatial scales, appearing extended along the continuum in the low-resolution map.   In the high-resolution map, the \ntdp{} emission is more scattered but remains visible around the envelope, extending beyond the \cont{} into a filamentary structure.   \dcn{} is also detected along the extended continuum emission at low resolution, while in the high-resolution maps it appears only marginally detected as scattered emission. The distribution of \dcop{} closely resembles that of \ntdp{} and \dcn{}. Visual inspection of the molecular and continuum emission at a resolution of $\sim$500~AU suggests that the inner envelope has a diameter of $\sim$1000--2400~AU, excluding the extended emission along the northeast–southwest direction.

\subsection{\ntdp{} Emission with Higher Spectral Resolution}\label{sec:N2Dp_Emission_higher_spectral_rsolution}
\begin{figure*}
    \centering
    \includegraphics[width=0.8\linewidth]{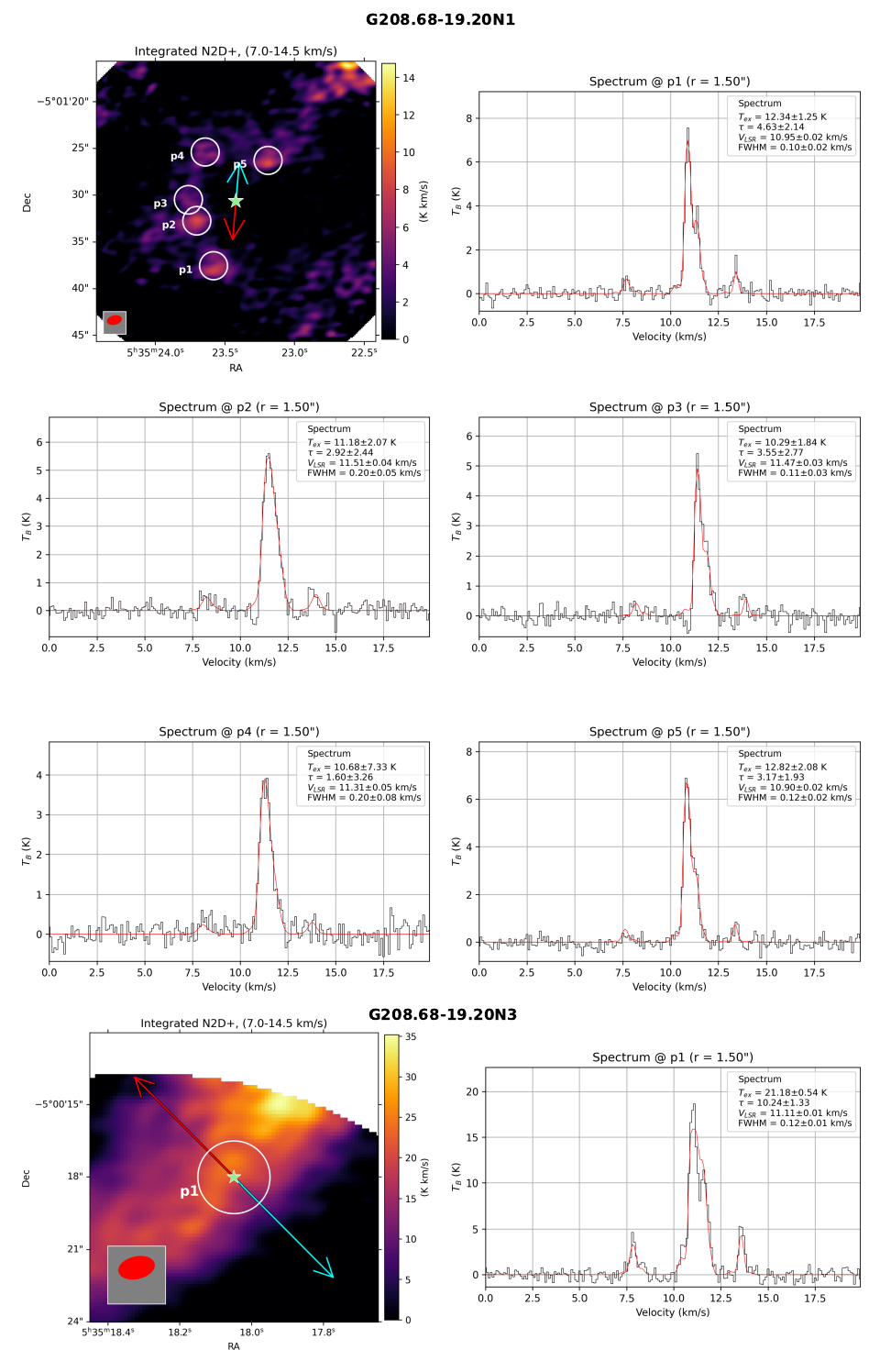}
\caption{\ntdp{} observed spectra (black) and corresponding model fits (red) for G208.68$-$19.20N1 and G208.68$-$19.20N3. 
Spectra are averaged over apertures with a radius of 1\farcs5, centered on positions p1, p2, \ldots\ as marked in the integrated maps (color images in the first panel of each object).  The derived parameters $T_{\mathrm{ex}}$, $\tau$, $V_{\mathrm{LSR}}$, and $V_{\mathrm{FWHM}}$ are indicated within each panel.  The continuum peak is marked with an asterisk, and the jet axes follow the orientation shown in Figure~\ref{fig:Appendix_COSiO_outflowJet_highresolution}.}
    \label{fig:n2dp_spectra_archive_data}
\end{figure*}

High spectral resolution observations of \ntdp{} at $\sim$ 0.1~\kms{} have been analyzed to investigate the kinematics and physical properties of the dense gas surrounding two cores, G208N1 and G208N3; these data do not cover G215M. The \ntdp{} line consists of numerous hyperfine (hf) components. Fitting these hyperfine lines allows us to derive key parameters such as the excitation temperature \(T_{\mathrm{ex}}\), optical depth \(\tau\), LSR velocity \(V_{\mathrm{LSR}}\), and line width \(V_{\mathrm{FWHM}}\). Figure~\ref{fig:n2dp_spectra_archive_data} shows the \ntdp{} spectra for G208N1 and G208N3. The spectra for G208N1 were extracted from regions with a radius of 1\farcs5 around the peak emission associated with the semi-arc-like envelope structure. Five such peaks, labeled p1–p5, were visually identified around the envelope (Figure \ref{fig:n2dp_spectra_archive_data} and Table \ref{tab:n2dp_fitting_params}). These regions correspond to the surrounding envelope and have peak velocities ranging from 10.3 to 11.5~km~s$^{-1}$, close to the systemic velocity, with a FWHM of only 0.1--0.2~km~s$^{-1}$. The velocity extends from 10 to 12.5~km~s$^{-1}$.  These locations therefore most likely unrelated to outflow emission, and the emission is not coincident with the outflow or jet axis. For G208N3, the spectrum extraction was performed at the core center, as marked in Figure~\ref{fig:n2dp_spectra_archive_data}. In all cases, we selected a radius of 1\farcs5 to ensure that at least one beam size of emission is included in the spectra. The spectra were fitted using the {\it pyspeckit} software package with the $n2dp$ model. The fitting parameters are overplotted as legends on the figure and listed in Table~\ref{tab:n2dp_fitting_params}.

G208N1 exhibits excitation temperatures \(T_{\mathrm{ex}}\) in the range 10--13~K, except for values from position P4, which show large uncertainties and poorly detected satellite lines. All these peaks were detected around the arc-like structure in the envelope, away from the continuum center. In contrast, \ntdp{} traces emission closer to the source in G208N3. Consequently, the excitation temperature (\(\sim 21~\mathrm{K}\)) is higher for G208N3 than for G208N1.

The optical depth in G208N3 (\(\sim 10.24 \pm 1.33\)) is a few times higher than in G208N1 (in a range \(\sim 2.9\)–4.0). The satellite groups are brighter in G208N3 than in G208N1, possibly because \ntdp{} in G208N1 has been depleted due to internal heating during its evolution, whereas G208N3 is younger and internal heating is insufficient to cause significant depletion. The LSR velocities are \(V_{\mathrm{LSR}} \sim 11.509~\mathrm{km\,s^{-1}}\) and \(11.134~\mathrm{km\,s^{-1}}\) for G208N1 and G208N3, respectively. The velocity line widths are similar for both sources. 

\subsection{Molecular Gas along the Outflows}\label{sec:outflow_emission}

\begin{figure*}
    \centering
    \includegraphics[width=0.85\linewidth]{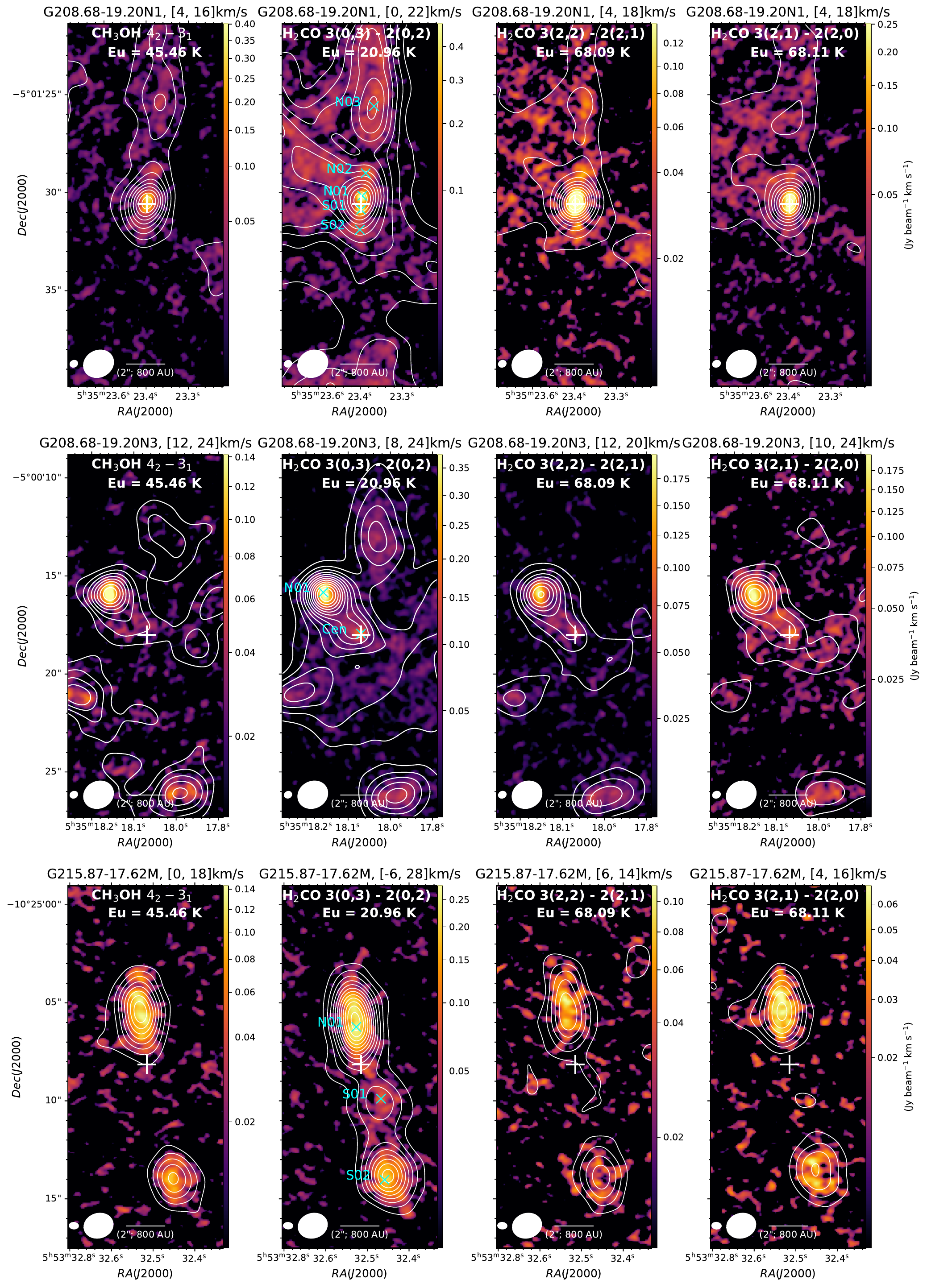}
    \caption{
High-resolution ($\sim$150~AU) emission maps integrated over the velocity ranges indicated in each panel of \chohlower{} (column 1), \htcoi{} (column 2), \htcoii{} (column 3), and \htcoiii{} (column 4). Rows 1, 2, and 3 correspond to the sources G208N1, G208N3, and G215M, respectively. White contours show low-resolution ($\sim$500~AU) taper maps integrated over the same velocity ranges, taken from Figure~\ref{fig:Appendix_multitracers_outflowJet_lowresolution_uvtap}, with contour levels starting at 3$\sigma$ and increasing in steps of 3$\sigma$. The $\sigma$ values (in mJy~beam$^{-1}$~km~s$^{-1}$) for each molecule and source are: G208N1 (0.030, 0.052, 0.035, 0.035), G208N3 (0.020, 0.045, 0.025, 0.025), and G215M (0.025, 0.035, 0.016, 0.016). The continuum peak is marked with an white plus ($+$).  Approximate knot centers are labeled N01, N02, ... in the northern lobe and S01, S02, ... in the southern lobe.
}
    \label{fig:youngest_JetsOutflows}
\end{figure*}

Figure~\ref{fig:youngest_JetsOutflows} presents \chohlower, \htcoi{}, \htcoii{}, and \htcoiii{} emission, which, together with CO (2--1) and SiO (5--4) (Figure~\ref{fig:Appendix_COSiO_outflowJet_highresolution}), trace collimated structures along the jet axis. All three sources exhibit similar velocity extents in CO and SiO, as well as comparable spatial morphology, which complicates the distinction between the collimated jet and the outflow \citep[see also][]{2024AJ....167...72D}. \ceo{} further traces the jet base near the sources (Figures~\ref{fig:EnvelopeEmission_highresolution}, \ref{fig:EnvelopeEmission_lowresolution}). In all three protostars, emission is detected along the jet axis. As shown earlier, higher \choh{} transitions peak in the envelope or jet base, while lower transitions such as \chohlower{} are more sensitive to extended jet/outflow emission. \htco{} behaves similarly, appearing both in the envelope and along the outflow.  

G208N1 (first row of Figure \ref{fig:youngest_JetsOutflows}) shows a unipolar jet/outflow extends northward. Low-resolution maps ($\sim$ 500 AU;  Figure \ref{fig:Appendix_multitracers_outflowJet_lowresolution_uvtap}), clearly show the outflow, while high-resolution data reveal \chohlower{} bright at the jet base and scattered near-source emission. The northern lobe extends beyond the \cto{} and \sio{} detections (Figure \ref{fig:Appendix_COSiO_outflowJet_highresolution}), implying a longer dynamical age. Both of these emission features exhibit similar velocity extent (velocity offset from –50 to +50 km s$^{-1}$) and comparable collimated spatial morphology \citep[][]{2024AJ....167...72D}. The lobe is also traced by \htcoi{}, which shows knots in the high-resolution map (Figure~\ref{fig:youngest_JetsOutflows}). Fainter emission from \htcoii{} and \htcoiii{} is detected along the lobe.  Unlike \ntdp{} emission (Section~\ref{sec:N2Dp_Emission_higher_spectral_rsolution}), H$_2$CO is mostly observed along the jet axis. Even near the source, within the inner envelope, it likely traces the base of the jet. The peak velocities range from 11--13~km~s$^{-1}$, although the emission extends from 0 to 22~km~s$^{-1}$, with a FWHM of 5~$\pm$~2~km~s$^{-1}$.

G208N3 (second row of Figure~\ref{fig:youngest_JetsOutflows}) shows that all four tracers peak at the tip of the northwest lobe, with weaker emission near the continuum. The morphology matches that of CO (2--1) and SiO (5--4) (Figure \ref{fig:Appendix_COSiO_outflowJet_highresolution}), and the compact extent indicates a short dynamical timescale. Comparable velocity extents (velocity offset from –66 to +58 km s$^{-1}$) and similar spatial morphology are observed for both CO and SiO emission \citep[][]{2024AJ....167...72D}.

G215M (third row of Figure~\ref{fig:youngest_JetsOutflows}) shows that \chohlower{} traces knots in both lobes, while \htcoi{} highlights both knots and diffuse emission across the outflow. The higher \htco{} transitions show a similar distribution to \chohlower{}, and the overall morphology agrees with CO (2--1) and SiO (5--4). Similar to G208N1 and G208N3, both CO and SiO emission features exhibit comparable extents (velocity offsets from –40 to +38 km s$^{-1}$) and collimated spatial morphology \citep[][]{2024AJ....167...72D}.

\subsubsection{Rotational Temperatures}\label{sec:H2COrotational_temperature} 
We have taken advantage of having three observed transitions of \htco{} to derive the rotational temperatures. As discussed earlier, the \htco{} morphology indicates that these lines could be tracing lower-density outflow material, higher-density knots, or may include contributions from envelope emission at the jet base. To determine the rotational temperatures, we measured the average intensities within a radius of 0\farcs5 around the peak position from the high-resolution ($\sim$150~AU) maps integrated over the emitting velocity range.  The centers of the selected regions corresponding to the knot peak positions of high-resolution maps are marked in second column of Figure~\ref{fig:youngest_JetsOutflows} (labeled N01, N02, …, S01, S02). The rotational diagrams for these regions are presented in Figure~\ref{fig:H2CO_rotational_diagram}, where each panel header indicates the corresponding region (N01, N02, …, S01, S02). These peaks positions are likely jet ``knots". The aperture radius ensures that at least one synthesized beam is included. The Einstein coefficients ($A_{ul}$) and upper-level energies ($E_u$) were adopted from the Splatalogue database\footnote{\url{https://splatalogue.online/}}, as listed in Table \ref{tab:Spectral_lines_used_in_this_work}.

 Assuming optically thin emission under local thermodynamic equilibrium (LTE) conditions, the column densities per statistical weight ($N^{\rm thin}_{\rm u}/g_{\rm u}$) are shown in Figure~\ref{fig:H2CO_rotational_diagram} as a function of the upper energy level ($E_{\rm u}$) of the lines. The column density of the upper level is given by  
\begin{equation}
N^{\rm thin}_{\rm u} = \frac{8\pi k \nu^2}{h c^3 A_{ul}}\, I ,
\end{equation}  
where the integrated line intensity is $I = \int T_B \, dv$, with $T_B$ being the brightness temperature. The best-fit rotational temperatures are indicated in each panel of Figure~\ref{fig:H2CO_rotational_diagram} and in Table \ref{tab:rotational_temperature}. Assuming these rotational temperatures to be representative of the excitation temperature of the jet/outflow, we have estimated the total column density, $N^{\rm tot}_{\rm H_2CO}$. 
 The 40\% flux uncertainty was adopted as a conservative estimate to account for additional sources of error beyond the nominal ALMA calibration uncertainty ($\sim$10\%), and typically flux uncertainty 10--20\% for most ALMA datasets. In our case, the spectra were extracted from relatively extended and scattered-intensity emission regions in a few cases (see Figure \ref{fig:youngest_JetsOutflows}), where uncertainties from baseline subtraction and missing short-spacing flux can contribute significantly. To avoid underestimating the propagated uncertainties in the derived quantities, such as rotational temperatures and column densities, we therefore adopted a total flux uncertainty of 40\%. The quoted errors on $T_{\rm rot}$ and $N_{\rm tot}$ are therefore the formal fitting errors derived from the covariance matrix of the weighted linear regression, given the assumed flux uncertainties.

The slopes of the emission near the continuum peak of G208N1 (positions N01 and S01) deviate from the expected trend in the rotation diagram, yielding unphysical (negative) temperature estimates; these values are therefore excluded. The anomalous slopes are likely due to optical depth effects, as the lower transition \htcoi{} appears optically thick near the source. Likewise, the \htcoii{} and \htcoiii{} emission at position S01 of G215M is weak or undetected and may be affected by optical depth, resulting in large uncertainties in the rotational temperature fitting (Table~\ref{tab:rotational_temperature}).

%

%

\subsubsection{\htco{} Non-LTE Analyses}\label{sec:H2COnonLTE} 

To investigate the opacity levels, we estimated the line opacities in the regions described above using a non-LTE radiative transfer approach. We employed the RADEX code \citep{2007A&A...468..627V}, following the methodology of \citet{2015ApJ...802..126H}, which utilizes an opacity-weighted radiation temperature to characterize the molecular excitation. 

RADEX computes the level populations of H$_2$CO by solving the statistical equilibrium equations, accounting for both collisional and radiative processes. This non-LTE treatment is essential in environments where the local density is insufficient to thermalize the molecular levels. The code applies an escape-probability formalism to approximate photon trapping, providing realistic estimates of line opacities without assuming LTE conditions.

\begin{figure*}
    \centering
    \includegraphics[width=0.8\linewidth]{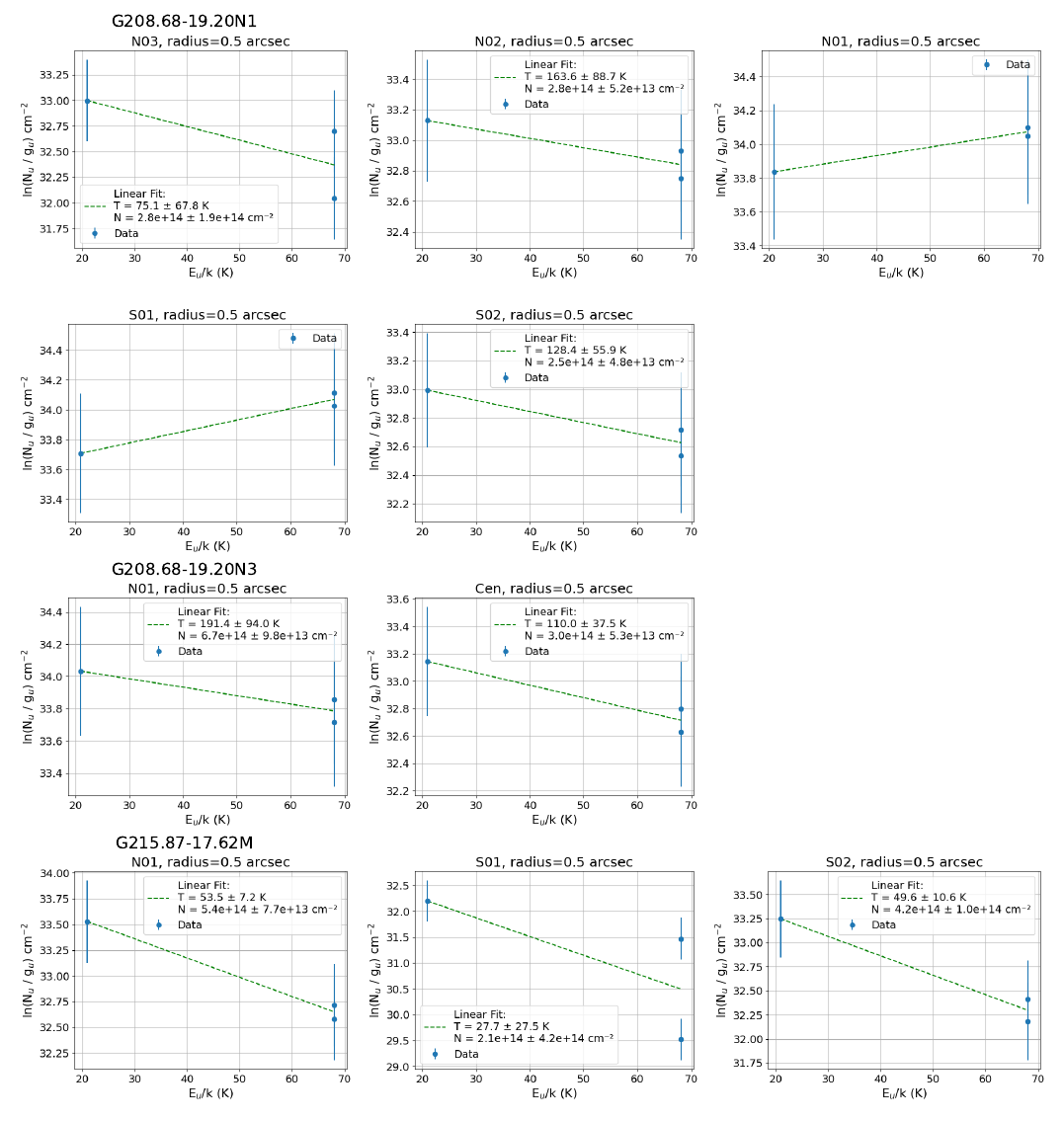} 
        \caption{Rotation diagrams for the molecular transitions of H$_2$CO are shown for three different objects. These diagrams are derived from the average intensities within 0\farcs5 along the outflow/jet axis, indicated by cross marks in Figure~\ref{fig:youngest_JetsOutflows}. The same regions around the knots, marked with circles, are labeled as N1, N2, \ldots, Cent, S1, S2. Error bars represent the assumed conservative uncertainties in our measurements, approximately 40\% of the estimated values. The green solid lines show linear fits to the observed data points. The rotational temperatures and column densities derived from these fits are displayed in the corresponding panels.
        }
    \label{fig:H2CO_rotational_diagram}
\end{figure*}

From our analysis, we find that the lower transitions of \htco{} (\htcoi{} and \htcoii{}) exhibit opacities of $\tau_\nu > 1.0$ in the N01 and S01 regions of G208N1, confirming that these transitions are optically thick, while the higher transition \htcoiii{} shows $\tau_\nu < 1.0$, indicating optically thin emission. This optical thickness explains the opposite slopes observed in Figure~\ref{fig:H2CO_rotational_diagram} (N01 and S01). The estimated fluxes of lower transitions in these regions represent lower limits, and the derived temperatures should therefore be regarded as upper limits. In contrast, all other regions in G208N1, G208N3, and G215M show $\tau_\nu < 1.0$, supporting the assumption of optically thin emission.


\begin{table}[ht]
\centering
\caption{Rotational temperatures and column densities of \htco{} for each peak in Figure \ref{fig:youngest_JetsOutflows}}
\begin{tabular}{lcc}
\hline

Region & $T_{\rm rot}$ (K) & $N_{\rm tot}$ (e+14\ cm$^{-2}$) \\
\hline
 & G208N1 & \\
N03 & $75.1 \pm 67.8$ & $2.83 \pm 1.94$ \\
N02 & $163.6 \pm 88.7$ & $2.77 \pm 0.52$ \\
N01 &      $--$        &       $--$         \\
S01 &     $--$         &     $--$         \\
S02 & $128.4 \pm 55.9$ & $2.51 \pm 0.48$ \\
\hline \hline
 & G208N3 & \\
N01 & $191.4 \pm 94.0$ & $6.73 \pm 0.98$ \\
Cen & $110.0 \pm 37.5$ & $3.00 \pm 0.53$ \\
\hline\hline
 & G215M & \\
N01 & $53.5 \pm 7.2$ & $5.39 \pm 0.77$ \\
S01 & $27.7 \pm 27.5$ & $2.05 \pm 4.18$ \\
S02 & $49.6 \pm 10.6$ & $4.19 \pm 1.02$ \\
\hline
\end{tabular}\label{tab:rotational_temperature}
\end{table}

\section{Discussion} \label{sec:discussion}

\begin{figure*}
    \centering
    \includegraphics[width=0.9\linewidth]{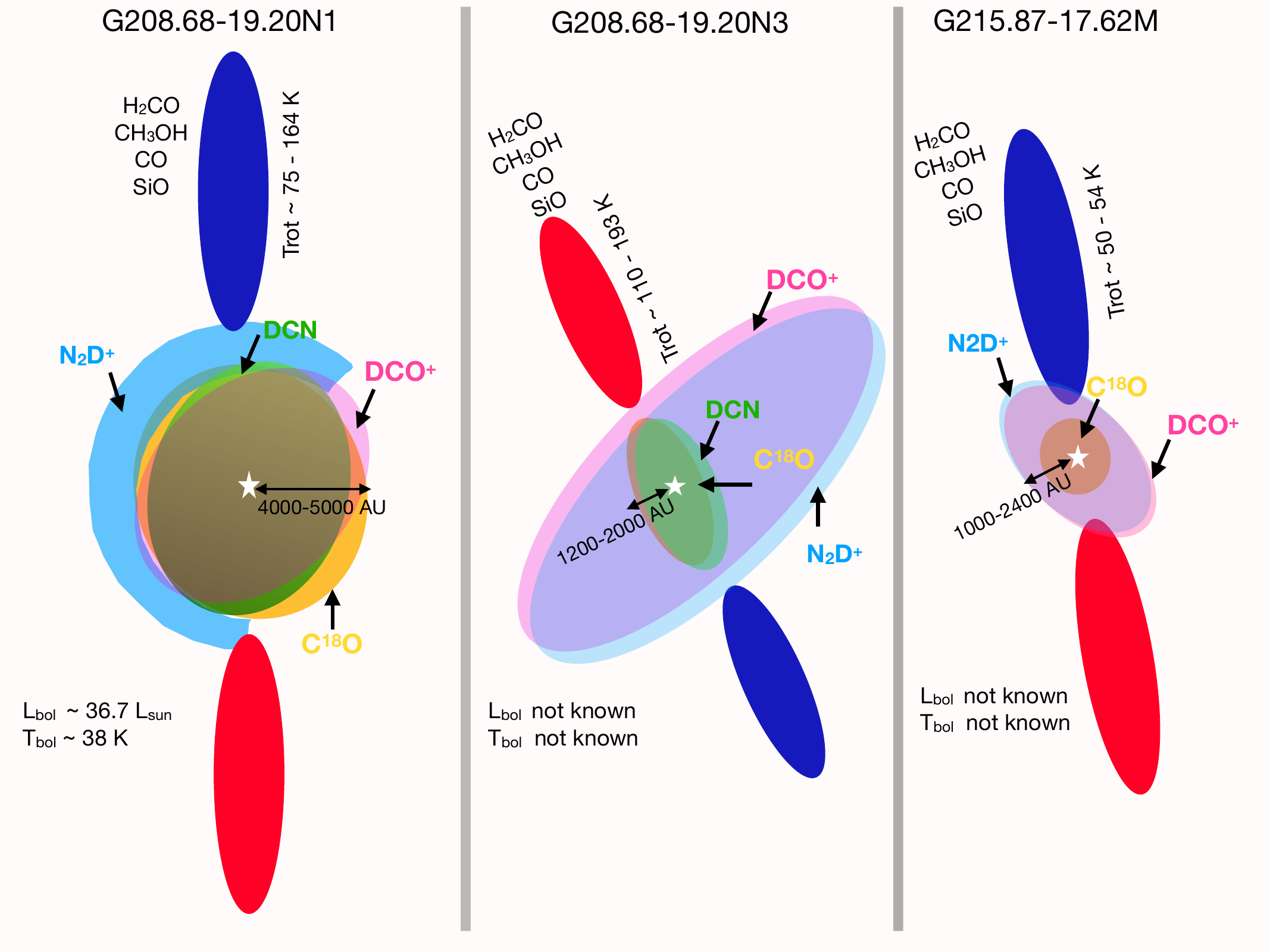}
    \caption{Cartoon summary of detected species for the three protostars. Molecular species are shown in distinct colors, scaled relative to their observed extents. Continuum peaks are marked by stars.}
    \label{fig:cartoon_summary}
\end{figure*}

The key observational results—including detections of multiple molecular species at a common angular resolution, their projected sizes, signatures of sublimation and depletion, and rotational temperatures measured along the jets—are summarized in Figure~\ref{fig:cartoon_summary} and Table~\ref{tab:summary_parameters}. These diagnostics collectively provide insight into the physical conditions and evolutionary stages of the protostellar cores.

\subsection{Local Gas Heating by the Protostars}
Radiation from the central protostar heats the surrounding medium. As the protostar evolves, its temperature and luminosity vary in response to both accretion and internal evolution, driving changes in the thermal and chemical structure of the envelope.

G208N1, previously classified as a hot corino \citep[]{2020ApJ...898..107H,2022ApJ...927..218H}, shows signatures of advanced chemical processing. The higher temperature in its inner envelope likely destroys \ntdp{} via reactions with CO, producing \dcop{}. Consequently, \ntdp{} is depleted near the continuum peak, forming an extended arc-like shell, while \ceo{} emission appears relatively faint. Its luminosity exceeds that of several benchmark Class\,0 sources e.g., HH~211, HH~111, B335 \citep[][]{2020A&ARv..28....1L,2024AJ....167...72D}, possibly due to a recent accretion burst in this source. Such a burst would shift the CO snowline outward and give rise to the large \ntdp{} shell surrounding the source. Notably, \ntdp{} exhibits excitation temperatures of $\sim$ 10--13 K in that ring (section~\ref{sec:N2Dp_Emission_higher_spectral_rsolution}). The detection of compact \chohhigher{} emission around the continuum peak and its absence across the extended envelope and the weak \ceo{} emission around the extended envelope are consistent with earlier findings \citep[][]{2024ApJ...976...29H}. 
These features likely result from freeze-out of CO and CH$_3$OH onto dust grains in the cold, dense outer envelope. Methanol is only released into the gas phase in the warm inner region near the protostar.
The enhanced \dcop{} toward the center implies that \htdp{} may still persist in this region. Detection of \dcn{} further indicates a warm environment, since it is formed both via warm channels involving \chtdp{} ($\sim 62$\%) and cold channels involving \htdp{} ($\sim 22$\%) \citep[]{2001ApJS..136..579T}.

H$_2$CO emission is primarily associated with the jet and outflows, as indicated by its absence in the inner core and its alignment along the outflow axis. Two bright knots near the jet base (Figure~\ref{fig:youngest_JetsOutflows}) suggest localized heating. H$_2$CO can form through gas-phase reactions (e.g., CH$_2$ + O, CH$_3$ + O) \citep[]{2017iace.book.....Y} or by CO hydrogenation on dust grains, followed by thermal desorption above $\sim$40~K \citep[]{2006A&A...457..927G}. The observed rotational diagrams for N01 and S01 region (Figure \ref{fig:H2CO_rotational_diagram}) indicate optical depth effects in the lower transitions (section \ref{sec:H2COnonLTE}). The detection of both C$^{18}$O and CH$_3$OH near the jet base supports their efficient release through sublimation, followed by subsequent gas-phase chemistry leading to the formation of H$_2$CO.

For G208N3, the envelope traced by \ntdp{} and \dcop{} is filamentary, while the core remains relatively compact in \dcop{}.  Interestingly, \ntdp{} exhibits excitation temperatures of $\sim$ 21.18$\pm$0.54 K around the protostar G208N3, which is significantly higher than that in ring like structure in G208N1 (section~\ref{sec:N2Dp_Emission_higher_spectral_rsolution}). Detection of \dcn{} and \dcop{} implies a moderately warm environment, though the weak \ceo{} and H$_2$CO suggest conditions close to the CO sublimation threshold. 

In G215M, high-resolution maps reveal compact \ceo{} emission confined to the continuum peak, while low-resolution maps show no extended component. \ntdp{} and \dcop{} are weak and irregular, and \dcn{} is absent. These signatures suggest a colder environment with inefficient synthesis of CO-derived species. 

 We note that the hyperfine temperatures derived from \ntdp{} (Table \ref{tab:n2dp_fitting_params}) are significantly lower than the rotational temperatures estimated from the \htco{} transitions (Table \ref{tab:rotational_temperature}). The \ntdp{} emission primarily traces cooler regions, mostly associated with the surrounding envelope, whereas the \htco{} transitions probe more excited regions in the jets or at the jet base within the envelope. Consequently, the \htco{} temperatures are naturally higher.

In summary, the relative sizes of \ceo{} emission, depletion of \ntdp{}, and presence of CO-derived molecules (\dcop{}, \dcn{}) collectively trace the evolutionary progression. G208N1 shows the most advanced chemistry and heating, G208N3 represents an intermediate stage, and G215M is likely the youngest source.

\subsection{Status of the Outflow Emission}

In G208N1, the SiO and CO morphology (Figure \ref{fig:Appendix_COSiO_outflowJet_highresolution}) and position--velocity diagrams \citep[][]{2024AJ....167...72D} show nearly identical velocity dispersions, complicating separation of outflow and jet components. The system shows a collimated jet with mean deprojected velocities $V_j \sim 102^{+24}_{-35}$ km\,s$^{-1}$ and a high mass-loss rate of $\dot{M}_j \sim 2.5 \times 10^{-6}\,M_\odot$\,yr$^{-1}$. The dynamical age is constrained to $\mathcal{T}_{\rm dyn} \sim 47^{+38}_{-10}$~yr based on compact SiO and CO emission, though the northern lobe likely extends farther, yielding a conservative estimate of $\sim$1000~yr.   H$_2$CO emission yields $T_{\rm rot} \sim 75$--164~K and $N_{\rm tot} \sim (2.51$--$2.83)\times 10^{14}$~cm$^{-2}$ in the knots (Figure~\ref{fig:youngest_JetsOutflows}; Table~\ref{tab:rotational_temperature}). Strong CH$_3$OH transitions further support high kinetic temperatures, consistent with CO sublimation and subsequent molecule formation along the jet.  

For G208N3, SiO and CO exhibit similar velocity dispersions, again making it difficult to separate jet from outflow. The kinematics suggest a young jet launched only a few hundred years ago, with a blue lobe dynamical age of $\mathcal{T}_{\rm dyn} \sim 228^{+160}_{-110}$~yr. The deprojected velocity is $V_j \sim 61^{+41}_{-15}$~km\,s$^{-1}$, and the mass-loss rate is $\dot{M}_j \sim 1.3 \times 10^{-6}\,M_\odot$\,yr$^{-1}$ (Table \ref{tab:summary_parameters}).   H$_2$CO shows high $T_{\rm rot} \sim 110$--193~K and $N_{\rm tot} \sim (3.0$--$6.7)\times 10^{14}$~cm$^{-2}$ in the outer knots, accompanied by CH$_3$OH emission with a similar spatial distribution. These results suggest that shock heating within the jet is sufficient to sublimate CO and form complex molecules, with minimal influence from protostellar UV radiation.  

In case of G215M, SiO and CO exhibit similar velocity structures, with a mean deprojected velocity of $V_j \sim 60^{+28}_{-13}$~km\,s$^{-1}$ and a mass-loss rate of $\dot{M}_j \sim 0.32 \times 10^{-6}\,M_\odot$\,yr$^{-1}$. The estimated dynamical age is $\mathcal{T}_{\rm dyn} \sim 292^{+133}_{-108}$~yr (Table \ref{tab:summary_parameters}).   The outflow chemistry indicates cooler conditions, with H$_2$CO rotational temperatures of only $T_{\rm rot} \sim 50$--54~K and column densities $N_{\rm tot} \sim (4.2$--$5.4)\times 10^{14}$~cm$^{-2}$, excluding the estimation at S01 with high errorbars. CH$_3$OH emission is weak and co-spatial with H$_2$CO, implying limited CO sublimation. 

\subsection{Evolutionary status of the Protostars}
G208N1 is a luminous Class\,0 protostar with a massive envelope and extended size of 4000-5000 AU (Table \ref{tab:summary_parameters}). Its very high luminosity indicates a recent outburst, which is supported by the \ntdp{} cavity in the continuum peak and semi-arc like structure surrounding the continuum. Its powerful, collimated jet exhibits high velocities and a large mass-loss rate, with a dynamical age possibly ranging from a few decades to $\sim$10$^3$~yr. Slightly elevated rotational temperatures of H$_2$CO and strong CH$_3$OH emission indicate efficient heating and CO sublimation along the jet, consistent with chemical enrichment driven by shocks. Together, these characteristics identify G208N1 as an energetic young protostar undergoing active accretion and feedback.  

In contrast to G208N1, G208N3 is less luminous and more poorly characterized, with a smaller envelope mass and smaller envelope size (1200-2000AU; Table \ref{tab:summary_parameters}). Its jet is younger and slower, a mass-loss rate about half that of G208N1, and a dynamical age of only a few hundred years. Despite its lower energetics, H$_2$CO and CH$_3$OH indicate similarly high rotational temperatures, consistent with shock heating and CO sublimation along the jet. These results suggest that while G208N3 drives a weaker and younger outflow than G208N1, the chemical enrichment processes in the jet are broadly comparable.

Compared to G208N1 and G208N3, G215M is less massive and less chemically evolved. With an envelope mass of only $\sim$0.3~$M_\odot$ and smaller size (1000-2400 AU; Table \ref{tab:summary_parameters}), it drives a slower and weaker jet with a dynamical age of a few hundred years. Unlike the hotter conditions in G208N1 and G208N3, the H$_2$CO rotational temperatures in G215M are relatively low, and CH$_3$OH emission is weak, suggesting limited CO sublimation and less efficient shock chemistry. These properties indicate that G215M is likely forming a low-mass star with reduced accretion and feedback, representing an earlier and less chemically enriched stage compared to the other two sources.  

It is important to note that the different morphologies of the species may arise from varying physical conditions, rather than being solely determined by protostellar evolution. For instance, if a source is in a quiescent accretion phase and thus has a low luminosity—as might be the case for G208N3 and G215M—the region where CO is not depleted would remain small, even if the source is relatively evolved. At the current angular resolution, such a compact region may not be detectable, which could lead to the source being misclassified as less evolved. 



\subsection{Comparison with Previous Studies of Low-Mass Protostars}
Several previous studies have demonstrated that molecular tracers provide powerful diagnostics of the thermal and chemical structure in deeply embedded protostars. Formaldehyde has long been used as a probe of gas temperatures in envelopes and outflows; for example, \citet{2004A&A...416..577M} and \citet{2005A&A...437..501J} showed that multi-transition H$_2$CO observations of envelopes in Class~0 sources at low-resolution using JCMT and IRAM telescope reveal abundance jumps when dust goes beyond grain evaporation limit, $\sim$ 100 K. Deuterated species, in contrast, are more sensitive to colder conditions. Surveys of N$_2$D$^+$ and deuterated formaldehyde toward young cores \citep{2007A&A...471..849R,2009A&A...493...89E} established that high deuteration fractions trace CO-depleted envelopes and decline with protostellar heating, making them robust evolutionary indicators.

High-resolution ALMA observations have since confirmed and extended this picture across several well-studied low-mass protostars. In the binary system IRAS~16293--2422, \citet{2018A&A...610A..54P} found H$_2$CO excitation temperatures of $\sim$100~K in the inner envelope, and formaldehyde forms in the ice as soon as CO has frozen onto the grains. In IRAS~15398--3359, \citet{2020ApJ...900...40O} showed that higher-energy H$_2$CO transitions are compact and warm near the source, whereas lower-energy lines are extended along the outflow, yielding excitation temperatures of 40--60~K in shocked regions.  Observations of VLA~1623--2417 as part of the FAUST program \citep{2025MNRAS.538.1481M} revealed strong spatial differentiation: hot-corino gas at $\sim$125~K, outflow cavities at 20--40~K, and cold streamers at $\leq$15~K with the highest deuteration levels. For L1551~IRS5, \citet{2020MNRAS.498L..87B} reported a hot corino around the protostellar binary with methanol and complex organics peaking at $\sim$100~K, confirming the existence of warm inner envelopes. More recently, FAUST observations of NGC~1333~IRAS~4A2 \citep{2025A&A...695A..78F} revealed compact CH$_3$OH and iCOM emission within 20--50~au, consistent with hot-corino chemistry, while radial temperature profiling demonstrated chemical segregation in the warm inner envelope.

 Together, these case studies highlight a consistent trend: in our targets, H$_2$CO transitions predominantly trace warm, shocked, or jet-base gas, while N$_2$D$^+$ and other deuterated tracers probe the colder envelope. We note, however, that H$_2$CO emission can also arise in a wider range of environments, including outflow cavities or streamers at lower temperatures ($<$20 K), as reported in studies of low-mass protostars. In our sources (G208N1, G208N3, and G215M), the elevated H$_2$CO excitation temperatures and spatial association with jet-like features suggest that the observed emission mainly traces warm, possibly shock-heated gas. This tracer-dependent thermal differentiation provides a robust framework for assessing the evolutionary stages of young protostars and situates our Orion sources within the broader context of well-studied low-mass systems.


\section{Summary and Conclusion} \label{sec:summary}
We have studied the inner envelopes of three protostellar cores to investigate the evolutionary structures at the very early stage of star formation. Based on the depletion and sublimation of various molecular species, their emitting sizes around the core, the outflow characteristics, we have drawn an evolutionary status of these three sources. 

(i) G208N1 envelope is detected with bright \ceo{}, \ntdp{} cavity, \dcn{}, \dcop{}, \chohhigher{}, \chohlower{}, \htco{}. The jet base and extended outflow emission are detected with \chohlower{} and \htco{} transitions. 

(ii) G208N3 envelope is detected with scattered extended \ceo{}, within filamentary \ntdp{} and \dcop{}, extended \dcn{} and \htco{} transitions. The outflow emission is detected with  \chohlower{} and \htco{} transitions.

(iii) G215M is detected with compact \ceo{}, scattered and weak  \ntdp{} and \dcop{}, extended \dcn{}. The extended outflow emission is detected with  \chohlower{} and \htco{} transitions.

(iv) The \htco{} rotational temperatures were estimated in the knots of the extended outflows for all three sources. Among them, G208N3 exhibits the highest $T_{\mathrm{rot}}$ values, while G215M shows the lowest. 

Based on their chemical compositions, sizes, envelope sublimation and depletion, continuum masses, and jet physical properties, it is evident that G208N1 (HOPS~87) is the most actively accreting and is likely the most evolved among the three protostars. G208N3 and G215M show weaker accretion activity; G208N3 may be younger than G208N1, whereas G215M is probably the youngest source. Variability in luminosity or accretion state could also influence their chemical morphologies. High-resolution studies to determine the protostellar central masses from envelope kinematics, along with sensitive multiwavelength observations to construct spectral energy distributions (SEDs), will help clarify the evolutionary distinctions among these sources. Expanding such chemical analyses to a larger sample of very young protostars will be crucial for statistically establishing chemical and morphological trends as diagnostic tools for understanding the earliest stages of protostellar evolution.


\section{Acknowledgments}.
\begin{acknowledgments}
We are thankful to the anonymous reviewer for their valuable suggestions, which helped to improve the overall quality of this paper.  We are thankful to Tien-Hao Hsieh for discussion at various stages. This paper makes use of the following ALMA data:  ADS/JAO.ALMA$\#$2018.1.00302.S. ALMA is a partnership of ESO (representing its member states), NSF (USA) and NINS (Japan), together with NRC (Canada), NSC and ASIAA (Taiwan), and KASI (Republic of Korea), in cooperation with the Republic of Chile. The Joint ALMA Observatory is operated by ESO, AUI/NRAO and NAOJ. C.-F.L. acknowledges the grant from the National Science and Technology Council of Taiwan (112-2112-M-001-039-MY3). This reserach is sponsored by NSTC 113-2124-M-001-008- and NSTC 114-2124-M-001-015-. D.J.\ is supported by NRC Canada and by an NSERC Discovery Grant. MJ acknowledges the support of the Research Council of Finland Grant 
No. 348342.  S.-Y.H. acknowledges support from the Academia Sinica of Taiwan (grant No. AS-PD-1142-M02-2).

\end{acknowledgments}

\vspace{5mm}
\facility{ALMA}\\
\software{Astropy \citep[][]{2013A&A...558A..33A,2018AJ....156..123A}, APLpy \citep[][]{2012ascl.soft08017R}, Matplotlib \citep[][]{2007CSE.....9...90H}, CASA \citep[][]{2007ASPC..376..127M}, pyspeckit \citep[][]{2022AJ....163..291G}}.\\

\appendix
\renewcommand{\thefigure}{A\arabic{figure}}
\setcounter{figure}{0}

\section{Outflow and Jets}\label{sec:Appendix_outflowjet}
\begin{figure}
    \centering
    \includegraphics[width=0.9\linewidth]{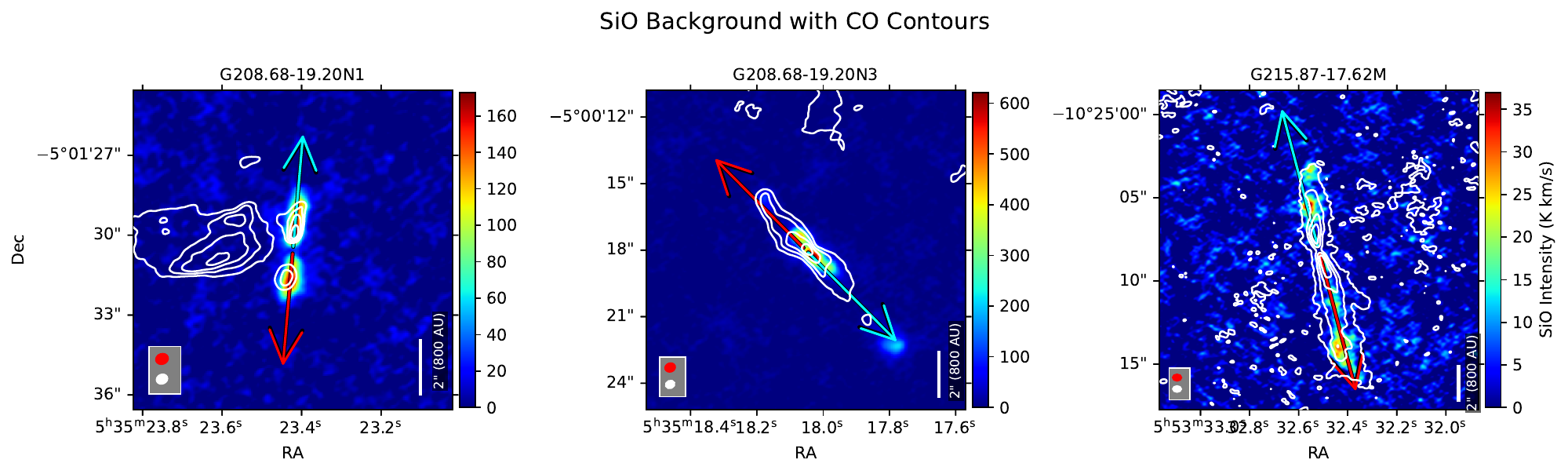}
    \caption{This image, reproduced following \citet[][]{2024AJ....167...72D}, illustrates the jet axis and compares the CO and SiO emission with the other molecular tracers presented in this study. SiO emission is shown in the background, with CO emission overlaid as contours. The contour levels span from the 95th percentile value of the CO data up to its maximum, divided into five intervals.  Beam sizes are indicated in the lower left corner, SiO in red and CO in white.  The blue- and red-shifted jet axes are marked with cyan and red arrows, respectively. }
    \label{fig:Appendix_COSiO_outflowJet_highresolution}
\end{figure}

SiO and CO emissions of G208N1, G208N3, and G215M are shown in Figure~\ref{fig:Appendix_COSiO_outflowJet_highresolution} at a resolution of $\sim$150~AU. In all cases, the emission traces narrow, collimated jets. We define the jet/outflow axis position angle (PA) as the angle measured east of north, with north at $0^\circ$ and east at $90^\circ$. A single PA in the range $0^\circ$ to $180^\circ$ is reported, corresponding to the direction of the blue-shifted lobe modulo $180^\circ$. For example, in G208.68$-$19.20N1, the blue-shifted lobe has a PA of approximately $355^\circ$, which corresponds to an outflow axis PA of $\sim175^\circ$. Similarly, the jets of G208N3 and G215M have PAs of $\sim45^\circ$ and $\sim15^\circ$, respectively.

\begin{figure}
    \centering
    \includegraphics[width=0.85\linewidth]{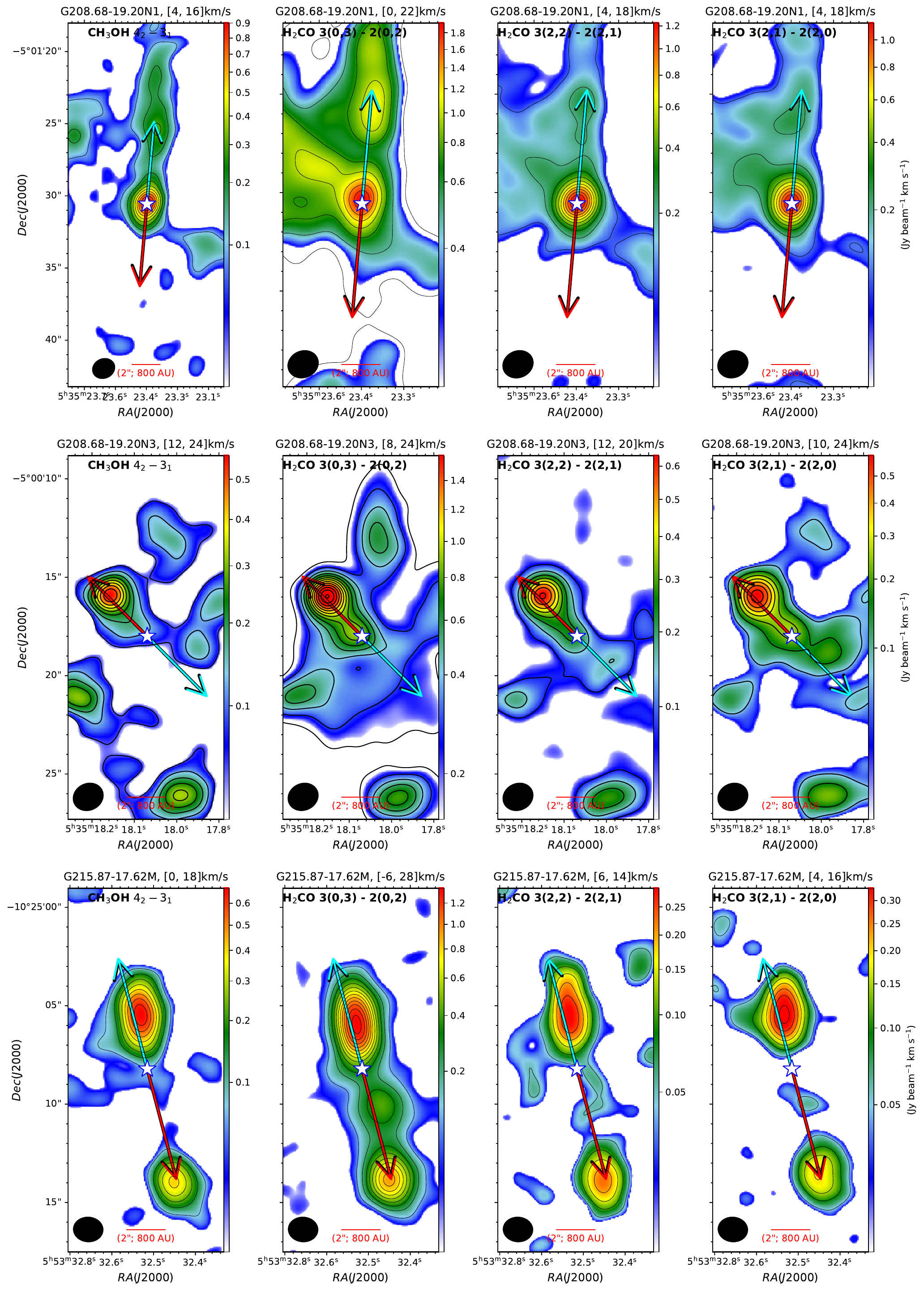}
\caption{Low-resolution UV-tapered ($\sim$500~AU) emission maps, integrated over the velocity ranges indicated in each panel, are shown for \chohlower{} (column~1), \htcoi{} (column~2), \htcoii{} (column~3), and \htcoiii{} (column~4). Rows correspond to the sources G208N1, G208N3, and G215M, respectively. Contours start at 3$\sigma$ and increase in steps of 3$\sigma$, where $\sigma$ (in mJy~beam$^{-1}$~km~s$^{-1}$) is: G208N1 — (0.030, 0.052, 0.035, 0.035); G208N3 — (0.020, 0.045, 0.025, 0.025); G215M — (0.025, 0.035, 0.016, 0.016). The continuum peak is marked with an asterisk, while blueshifted and redshifted jet axes are indicated by cyan and red arrows, respectively. Scalebars are shown, and beam sizes are indicated in the lower-left corner of each panel.}
    \label{fig:Appendix_multitracers_outflowJet_lowresolution_uvtap}
\end{figure}

\choh{} and \htco{} Emission along jet axis at low resolution ($\sim$ 500 AU) are shown in Figure \ref{fig:Appendix_multitracers_outflowJet_lowresolution_uvtap}.

\bibliography{sample631}{}
\bibliographystyle{aasjournal}

\end{document}